\documentstyle[aps,prd,eqsecnum,preprint,tighten,floats,epsf]{revtex}

\begin{document}

\preprint{UFIFT-HET-05-15}

\date{July 25, 2005}

\title{Robustness of Discrete Flows and Caustics \\ 
in Cold Dark Matter Cosmology}

\author{Aravind Natarajan and Pierre Sikivie}

\address{
   Institute for Fundamental Theory,\\	
   Department of Physics,\\
   University of Florida,\\
   Gainesville, FL,  32611-8440\\}

\maketitle

\begin{abstract} 

Although a simple argument implies that the distribution of dark matter 
in galactic halos is characterized by discrete flows and caustics, their
presence is often ignored in discussions of galactic dynamics and of
dark matter detection strategies.  Discrete flows and caustics can in 
fact be irrelevant if the number of flows is very large.  We estimate the
number of dark matter flows as a function of galactocentric distance and
consider the various ways in which that number can be increased, in
particular by the presence of structure on small scales (dark matter
clumps) and the scattering of the flows by inhomogeneities in the matter
distribution.  We find that, when all complicating factors are taken into 
account, discrete flows and caustics in galactic halos remain a robust
prediction of cold dark matter cosmology with extensive implications for
observation and experiment.

\vspace{0.34cm}
\noindent
PACS number: 98.80 Cq
\end{abstract}

\newpage

\section{Introduction and Outline}

A central problem in dark matter studies today \cite{CDM} is the question
of how the dark matter is distributed in the halos of galaxies, and in
particular in the halos of spiral galaxies such as our own Milky Way.  
Indeed, knowledge of this distribution is of crucial importance in trying
to understand galactic dynamics and in predicting signals for direct and
indirect searches for dark matter on Earth.  In this paper, we argue on
general grounds that the distribution of cold collisionless dark matter
(CDM) in galactic halos is characterized by discrete flows and caustics
\cite{ips,stw,crdm,sin,Tre}.  The latter are found to be a robust
prediction of the cold dark matter hypothesis without any further
assumptions, whether of symmetry, self-similarity or anything else.

In Section II we give the basic argument for the existence of discrete
flows and caustics, which is that cold dark matter particles lie at all
times on a 3-dim. hypersurface in 6-dim. phase space \cite{ips}.  We call
this 3-dim. hypersurface the ``phase space sheet".  The number of
discrete flows at a given physical location is the number of times this
sheet covers 3-dim. physical space at that location.  At the boundaries
between regions in physical space with differing numbers of flows, the
phase space sheet is tangent to velocity space.  As a result, the dark
matter density is very large on these surfaces, which are called
caustics.  The density diverges at the caustics in the limit of zero
velocity dispersion.

There can be no real doubt that CDM particles form discrete flows
and caustics for the reason just stated.  However, it is possible 
that this is not relevant in practice because the number of flows 
is so large that they form effectively a 6-dim. phase space 
continuum.  In Sections III to V, we estimate the number of flows 
and consider the ways in which this number can become very large.  

In Section III, we estimate the number of flows when the presence
of structure in the primordial dark matter fluid on scales less 
than galactic scales is neglected and scattering of the dark matter 
flows by inhomogeneities in the galaxy is neglected.  We provide a 
formula for estimating the number of flows at an arbitrary location 
in a galactic halo in these idealized circumstances.  At our location 
in the Milky Way halo the number of flows is of order 100.

In Section IV, we consider the effect of small scale structure 
in the dark matter fluid.  We show that small scale structure 
gives an effective velocity dispersion $\delta v_{\rm eff}$ to 
the flows.  However $\delta v_{\rm eff}$ is expected to be far 
less than what is required to smear out the 100 flows at our location.

In Section V, we consider the effect on the flows of dark matter 
of gravitational scattering by inhomogeneities in the galactic 
matter distribution.  We show that the known inhomogeneities in
the luminous matter of our galaxy are far insufficient 
to thermalize the flows of particles that have fallen in and 
out of the Galaxy less than 10 times in the past, implying that 
there are at least 20 flows at our location in the Milky Way which 
are not thermalized.  We also consider the effect of inhomogeneities 
in the dark matter distribution.  They can diffuse the remaining 
flows of dark matter only if a fraction of order one of the dark 
matter is in clumps of mass $10^{10}~M_\odot$ or more. 

In Section VI, we point to the existence of ripples in the distribution 
of light around large elliptical galaxies \cite{Mal,Her} as proof that
flows do not get diffused by gravitational scattering, or by anything 
else for that matter.  In particular the ripples are inconsistent with 
the dark matter being in clumps of mass $10^{10}~M_\odot$ or more.

In Section VII, we consider the effect of our galactic neighbor M31 
on the flows and caustics in the Milky Way.  We also discuss the 
ability of N-body simulations to resolve dark matter flows and 
caustics.

Finally, in Section VIII, we consider the implications of dark 
matter flows and caustics for experiment and observation.

\section{The basic argument}

Consider the phase space distribution of CDM particles at an early time
$t_{\rm in}$, long before density perturbations have become non-linear.  
Because CDM is `cold' all the dark matter particles at the same physical
location $\vec{q}$ have the same velocity $\vec{v}(\vec{q})$ up to a 
small dispersion $\delta v$, hereafter called the ``primordial velocity 
dispersion".  The primordial velocity dispersion of the leading cold 
dark matter candidates is very small.  For axions, it is \cite{cha}
\begin{equation}
\delta v_a(t) \sim 3 \cdot 10^{-17}~c~
\left({10^{-5} {\rm eV} \over m_a}\right)^{5 \over 6}
\left({t_0 \over t}\right)^{2 \over 3}~~~\ ,
\label{dva}
\end{equation}
whereas for weakly interacting massive particles (WIMPs)
\begin{equation}
\delta v_W(t) \sim 10^{-11}~c~
\left({{\rm GeV} \over m_W}\right)^{1 \over 2}
\left({t_0 \over t}\right)^{2 \over 3}~~~\ .
\label{dvw}
\end{equation}
Here $m_a$ and $m_W$ are respectively the axion mass and WIMP mass, 
$t$ is the age of the universe and $t_0$ is its present age.  Eqs.
(\ref{dva},\ref{dvw}) assume that $t$ is after the time of equal energy
density in matter and radiation. Since the particles at the same physical
location have the same velocity, they lie on a 3-dim. hypersurface in
6-dim. phase space \cite{ips,Tre}.  Let us call this hypersurface the
``phase space sheet".  The thickness of the sheet is $\delta v$.  The 
sheet is continuous because the density of particles is huge on the 
scale over which the sheet is bent in phase space.  Being collisionless, 
the particles move under the influence of purely gravitational forces.  
At a later time $t$, all the particles that were initially at position
$\vec{q}$ have moved to position $\vec{x}(\vec{q},t)$, where they have
velocity $\vec{v}(\vec{q},t) = {\partial \vec{x} \over \partial
t}(\vec{q},t)$.  The function $\vec{x}(\vec{q},t)$ therefore determines
the position of the phase space sheet at all times.

When the density perturbations become non-linear the sheet begins to 
fold in phase space, i.e. it begins to cover physical space multiple 
times whereas it covered physical space only once when density perturbations 
were small.  Mathematically this is expressed by stating that, at late 
times $t$, for a given physical location $\vec{r}$ there are in general
multiple solutions $\vec{q}_j$, $j = 1, 2~..~n$, to the equation 
$\vec{r} = \vec{x}(\vec{q}, t)$.  Each solution corresponds to a
distinct flow  at $\vec{r}$ with velocity 
$\vec{v}_j(\vec{r},t) = {\partial \vec{x} \over \partial t}(\vec{q}_j,t)$
and density  
\begin{equation}
d_j(\vec{r}, t) = 
{d(\vec{q}_j, t_{\rm in}) \over |D(\vec{q}_j, t)|}~~~~\ ,
\label{dmden}
\end{equation}
where $d(\vec{q}_j, t_{\rm in})$ is the density at location $\vec{q}_j$
at the early time $t_{\rm in}$, and 
\begin{equation}
D(\vec{q}, t) =
det \left({\partial x_k \over \partial q_l} \right)~~~\ .
\label{jacob}
\end{equation}
The magnitude of $D$ is the Jacobian of the map $\vec{q} \rightarrow
\vec{x}$.  

Caustics occur where $D=0$, {\it i.e.} where the map is singular
\cite{sin}.  In particular, there is a caustic wherever the number 
$n$ of flows changes.  The physical density is very large at the 
caustic because the phase space sheet is tangent to velocity space 
there.  The density diverges at the caustic in the limit of zero 
velocity dispersion.  Since the map $\vec{q} \rightarrow \vec{x}$ is
singular where the number of flows changes, caustics lie generically 
at the boundaries between regions which have different numbers of 
flows.  On one side of a caustic surface are two more flows than 
on the other.

The existence of discrete flows and caustics was noticed in past 
investigations.  In particular, caustics appear as simple fold
catastrophes in the distribution of dark matter on very large 
scales in the context of the Zel'dovich approximation \cite{Zel}, 
and in the distribution of dark matter on galactic scales in the 
context of the self-similar model of Filmore and Goldreich \cite{Fil}, 
and of Bertschinger \cite{Ber}.  That these investigations involve 
some approximations may have led to the mistaken belief, sometimes 
expressed, that discrete flows and caustics are a consequence of 
the approximations made rather than of just the collisionless dark 
matter hypothesis, i.e. that the discrete flows and caustics disappear 
when one goes beyond the Zel'dovich approximation in the treatment of 
density perturbations, or when one abandons the assumptions of spherical
symmetry and self-similarity in modeling galactic halos.   

However, as the argument given above shows, the existence of discrete 
flows and caustics follows from just the hypothesis of cold collisionless
dark matter.  No other assumptions are needed.

It is possible, nonetheless, that discrete flows have little relevance 
in practice.  The central reason for the practical irrelevance of 
discrete flows and caustics would have to be that the number of flows 
is very large, i.e. that the phase space sheet covers physical space so
many times that, for all practical purposes, it forms a 6 dim. continuum.  
We will address this issue now by estimating the number of flows and
considering all concievable ways in which that number can become large.

\section{How many flows?}

First, let us estimate the number of flows at an arbitrary location in 
a galactic halo as a function of galactocentric distance.  This first
estimate \cite{ips} is made under idealized conditions where we neglect 
the formation of structure on scales smaller than that of the galaxy as a
whole, as well as the diffusion of the flows by gravitational scattering 
off inhomogeneities in the galaxy.  After estimating the number of flows in
these idealized conditions we will consider the effect of the complicating
factors.

The number of flows at a given location inside a galactic halo is the
number of ways dark matter particles can reach that location from the 
far past.  If the gravitational potential is smooth on the scale ($\sim$
100 kpc) of the galaxy, the number of ways particles can reach a given
location from the far past is limited by, and of order, the number of
oscillations through the galactic potential that particles at that
location may have had since $t=0$.  Let us explain.

Very far from the galactic center, say at $r \sim$ 1 Mpc in the case of
our own Milky Way, there is only one flow because particles can reach
such a location from the far past in only one way, namely by falling
there on the way to the galactic center.  We are for the time being
ignoring the presence of our galactic neighbor M31, but will consider
later any role it may have.  At somewhat smaller distances from the
galactic center, there are three flows, corresponding to three ways such
a position can be reached from the far past.  Qualitatively the three
ways are 1) by falling there while on the way to the galactic center for
the first time, 2) by falling through the galaxy from the opposite side
and then reaching the position under consideration on the way out, and 3)
by falling through the galaxy from the opposite side, going all the way
out on the same side as the position under consideration, and then
reaching that position on the way back in.

Note that we can be absolutely certain that the number of flows changes
from one to three at some point when approaching the galaxy because the
number of solutions $\vec{q}_j$ of the equation $\vec{r} =
\vec{x}(\vec{q},t)$ increases when $r = |\vec{r}|$ decreases and the
number of solutions can only change by two at a time.  We are also
absolutely certain that there is a caustic at the boundary between the
region with one flow and the region closer to the galactic center with
three flows.  Indeed the map $\vec{q} \rightarrow \vec{x}$ is singular 
on that boundary and hence the density diverges there in the limit of 
zero velocity dispersion.

Further in there is a region with five flows.  For every point in that
region there are five ways in which particles can reach that given
location from the far past.  The first three are the same as 1) - 3)  
above. The additional two are (qualitatively):  4) by falling to the
galactic center from the same side as the location under consideration,
falling out on the opposite side, falling back in on the opposite side,
and now going past the location while falling out of the galaxy for the
second time, and 5) by falling in from the same side as the location
under consideration, falling out on the opposite side, falling back in
from the opposite side, falling out for the second time, turning around
and now going past the location while falling onto the galaxy for the
third time.  Still closer to the galactic center is a region with seven
flows, then a region with nine flows, and so on.  At each boundary where
the number of flows increases by two on the way in, there is a caustic.  
These caustics are simple fold catastrophes located on topological
spheres surrounding the galaxy.  We call those the ``outer caustics" of
the galaxy.  In addition there are ``inner caustics", as discussed in
refs. \cite{stw,crdm,sin,nat}.

Let us consider the region where there are seven flows.  The particles 
there oscillated through the galaxy up to three times. In this discussion, 
we are calling ``one oscillation"  the motion by which a particle starts
with zero radial velocity at some galactocentric distance $r$, approaches
the galactic center and then moves back out to a galactocentric distance
of order $r$.  The reason there are less than nine flows in this region
is that to produce nine flows some particles would have to oscillate four
times, and this would take longer than the age of the universe at that
distance from the galactic center.  Closer to the galactic center there
are nine flows because the oscillation time there is less than
approximately one fourth the age of the universe. Still closer to 
the galactic center there are eleven flows because the oscillation 
time there is less than one fifth the age of the universe.  And so on.
So we may estimate the number of flows $n(r)$ at galactocentric distance
$r$ by the formula 
\begin{equation} 
n(r) \sim 2~{t_0 \over T(r)} 
\label{nr} 
\end{equation} 
where $T(r)$ is the oscillation period through the galaxy with amplitude
$r$.  Eq. (\ref{nr}) is only valid in order of magnitude because it does
not take account of the fact that the galactic potential, and hence the
oscillation period, is time dependent.  

To improve on this, let us assume that the galactic potential is such
that the rotation curve is flat with present rotation velocity 
$v_{\rm rot}(t_0)$ and that in the past the rotation velocity 
increased according to the power law 
$v_{\rm rot}(t) = v_{\rm rot}(t_0)({t \over t_0})^p$.  It is easy 
to show that in such a potential, the product $r~v_{\rm rot}$ is an
adiabatic invariant, and that the oscillation period of a particle
decreases in time as $T \propto t^{-2p}$.  The improved version of
Eq. (\ref{nr}) is then found to be
\begin{equation} 
n(r) \simeq {2 t_0 \over (2p+1) T(r,t_0)} 
\label{nri} 
\end{equation}
where $T(r,t_0)$ is the present oscillation period with amplitude $r$.
At our distance, $r_\odot \simeq$ 8.5 kpc, from the center of the Milky 
Way the oscillation period is approximately 
$T(r_\odot,t_0) \simeq 1.3 \cdot 10^8$ years.  Since 
$t_0 \simeq 1.4 \cdot 10^{10}$ years, the number of 
flows at our location in the Galaxy $n_\odot \simeq 215/(2p+1)$.  The
exponent $p$ is related to the parameter $\epsilon$ \cite{Fil} of the
self-similar infall model by: $p = {2 \over 9\epsilon} - {1 \over 3}$.  
A favored range for $\epsilon$ is 0.2 to 0.35 \cite{stw}, which yields
$n_\odot$ in the range 84 to 134.  In conclusion, we expect the number 
of flows at the Sun's position in the Milky Way halo to be of order 100. 
This estimate should be valid within a factor 2 at least.

We estimate of order 100 flows on Earth in the idealized situation 
where the gravitational potential is smooth on galaxy scales and 
where the infalling dark matter has no small scale structure of its
own.  We will consider below the effect of small scale structure in 
the gravitational potential and in the infalling dark mattter, both of
which tend to increase the number of flows.

Before we get there, however, let us ask whether any effect can reduce the
number of flows below the above estimate.  In particular, it may seem that
angular momentum keeps some of the infalling particles from reaching us
since angular momentum introduces for each particle a distance of closest
approach to the galactic center.  However, angular momentum does not in
fact reduce the number of flows.  To see this, note that angular momentum
is a vector field tangent to the turnaround sphere of particles which are
about to fall in. A vector field on a topological sphere must have at
least two (simple) zeros.  That means that every infalling shell includes 
particles which will pass through the galactic center.  In fact, one can
show - see for example the discussion in ref. \cite{crdm} - that the
continual infall of particles from all directions in and out of a 
gravitational potential produces at least two flows at every point 
which is inside both the initial and final turnaround spheres.  The 
only way the particles would fail to reach such a point is for it to 
be near the top of a potential barrier.  However, because gravity is 
attractive, the gravitational potential does not have any maxima.

\section{Small scale structure in the dark matter fluid}

In the above description we have assumed that the dark matter 
falling onto a galaxy is without structure of its own.  In CDM 
cosmology, however, the spectrum of primordial density perturbations 
has power on all scales.  So the dark matter that is falling onto 
a galaxy has in general clustered previously on smaller scales.  
What effect does that have?  We show in this section that, provided 
the infalling clumps are not too large, the description of a dark 
matter halo in terms of a phase space sheet, discrete flows and 
caustics is still valid \cite{crdm,sin}.  It is of course understood 
here that the flows exist now in an average sense (i.e. after averaging 
over the clumps).

Where a clump has formed the sheet is wound up in phase space on the 
scale of the clump.  The sheet has acquired at that location a number of
sublayers which give it a certain thickness.  It describes therefore a
flow with an effective velocity dispersion $\delta v_{\rm eff}$.  $\delta
v_{\rm eff}$ is equal to the velocity dispersion of the particles in the
clump.  From the point of view of an observer located at some point in
physical space, the clumpiness of a flow passing by means that instead of
a single flow with a unique velocity there is a set of subflows, odd in
number, all with the same velocity up to the dispersion $\delta v_{\rm
eff}$.  $\delta v_{\rm eff}$ generally varies with location on the phase
space sheet since the size of the clumps that make up the sheet generally
varies with location.  Also the number of sheet sublayers varies with
location for the same reason.

The question is: how large can the effective velocity dispersion become
before the phase space structure described earlier loses its meaning?  
The answer is that the phase space sheet should not become so thick that
successive layers of the sheet touch one another and thus lose their
identity.  Where the different layers of the sheet touch, the flows
overlap in velocity space and become confused with one another.  So let
us ask: how large can the effective velocity dispersion be before the
approximately 100 flows at the Earth's location in the Galaxy become
confused with one another.  That question is simple to answer because 
the velocities are all in a 3-dim box in velocity space whose size is
determined by the escape velocity from the Galaxy at our position.  That
escape velocity is of order 600 km/s.  So the box is, roughly speaking, 
a cube of size 1200 km/s.  There is no reason for the 100 velocities to
cluster in a particular region of this cube.  If $\delta v_{\rm eff}$ is
equal or less than say 30 km/s, then clearly many of the flows on Earth
will be distinct from one another.  Now, it is very unlikely that the
clumps of infalling dark matter have velocity dispersion as large as 30
km/s.  Indeed 30 km/s is of order the velocity dispersion of the large
magellanic cloud (LMC).  If the infalling dark matter is in clumps as
large as the LMC, one would expect the clumps to be luminous, as is the
LMC.

In summary, provided the effective velocity dispersion of the dark matter
falling onto the galaxy is small enough (30 km/s or less in the case of
the Milky Way), the phase space structure of galactic halos is still
characterized by discrete flows and caustics.  However each flow may 
be divided into an odd number of subflows with velocity spread 
$\delta v_{\rm eff}$.  Likewise each caustic may be spread into 
an odd number of subcaustics, which are identical in shape but 
displaced relative to one another by distances proportional to 
$\delta v_{\rm eff}$.

\section{Gravitational scattering by galactic inhomogeneities}

When a dark matter flow passes by a clump of matter, whether dark 
matter or baryonic matter, the particles in the flow are scattered by 
the gravitational potential of the clump.  Examples of clumps of baryonic
matter include stars, globular clusters and molecular clouds.  For every
flow upstream of the clump, there are three flows downstream \cite{wic}.  
Let us  call them daughters of the original flow. Downstream of two clumps
there will be nine granddaughter flows, and after $p$ clumps, the number
of descendant flows will be $3^p$.  Since the number of descendants
is an exponentially growing function of the number of clumps, the total
number of flows is huge.  This would appear to destroy any notions of
phase space sheet, discrete flows and caustics.  But it doesn't. The
reason is that not all descendants are equal.  Although there are three
daughter flows downstream of a clump for every flow upstream, there is
very little density in the daughter flows except where their velocity
vectors have the same direction as the original flow.

Consider a flow of particles passing through a region populated 
by a class of objects of mass $M$ and density $n$.  Gravitational
scattering by the objects causes the velocity of each particle in 
the flow to have a random walk in velocity space.  This results in 
a diffusion of the flow over a cone of angle $\Delta \theta$.  One 
readily finds that \cite{ips}
\begin{eqnarray}
(\Delta \theta)^2 &=& 
\int~dt~\int_{b_{\rm min}}^{b_{\rm max}}~
{4 G^2 M^2 \over b^2 v^4} n v 2 \pi b~db \nonumber\\
&=& 1.8 \cdot 10^{-7}~\left({10^{-3} c \over v}\right)^3~
\left({M \over M_\odot}\right)^2~\ln ({b_{\rm max} \over b_{\rm min}})~
\left({t \over 10^{10} {\rm year}}\right)~
\left({n \over {\rm pc}^{-3}}\right)~~~~\ ,
\label{scat}
\end{eqnarray} 
where $v$ is the velocity of the flow and $t$ is the time over
which it encountered the objects in question.  In the galactic 
disk, giant molecular clouds are most likely the main contributors.
With $M \sim 10^6~M_\odot$, $n \sim 3/{\rm kpc}^3$, $b_{\rm max} \sim$ kpc
and $b_{\rm min} \sim 20$ pc, they yield $\Delta \theta \sim 0.05$
for dark matter particles that have spent most of their past in the 
galactic disk.  The contributions from globular clusters 
($M \sim 5 \cdot 10^5~M_\odot$, $n \sim 0.3/{\rm kpc^3}$) and
stars ($M \sim M_\odot$, $n \sim 0.1/{\rm pc}^3$) are less important.
Therefore, flows of dark matter particles that have spent most 
of their past in the central parts of the Galaxy may well be
washed out by scattering.

However, as was emphasized above, there are in the halo flows of particles
which are falling in and out of the galaxy for the first time as well as
flows of particles which have fallen in and out of the galaxy only a small
number of times in the past.  Since the particles involved have spent very
little of their past in the inner parts of the galaxy, such flows are not
washed out by scattering off molecular clouds or other inhomogeneities
in the luminous matter.  To be specific, consider particles which have 
fallen through the inner parts of the Galaxy ten times or less in the 
past.  Such particles spent at most $5 \cdot 10^8$ years in the disk, 
and hence $\Delta \theta < 10^{-2}$ for the corresponding 20 flows.  
Hence there are at least 20 dark matter flows on Earth which have not 
been thermalized by inhomogeneities in the distribution of luminous matter.
 
We must still consider the effect of scattering by inhomogeneities 
in the dark matter.  For a flow of particles falling in and out 
of the galaxy the dominant contribution to the integral $\int n dt$ 
in Eq. (\ref{scat}) is from the inner regions of the halo because 
$n$ decreases with $r$, at large $r$, as $r^{-2}$ or faster.  Let us
assume that a fraction $f$ of the dark matter is in clumps of mass $M$.  
Then, for a flow of velocity $v = 400$ km/s, passing through the inner 
parts (say $r <$ 20 kpc) of the Galaxy once, the effect of scattering 
by dark matter clumps is 
\begin{equation}
(\Delta \theta)^2 \sim 10^{-11}~f~\left({M \over M_\odot}\right)
\label{dmc}
\end{equation} 
where we used $M~n~=~f~{1 \over 3}~10^{-24}$ gr/cm$^3$ and 
$\ln({b_{\rm max} \over b_{\rm min}}) = 4$.  Unless a large 
fraction of the dark matter is in clumps of mass $10^{10}~M_\odot$ 
or larger, the flows of particles which have fallen in and out of 
the galaxy only a small number of times in the past are not thermalized.
We do not know of any evidence that the dark matter is in clumps of 
$10^{10}~M_\odot$ or larger.  To the contrary, the next subsection 
presents evidence against this possibility.

\section{Ripples around giant elliptical galaxies}

Malin and Carter \cite{Mal} observed that many bright elliptical galaxies
are surrounded by ripples in the distribution of light.  These ripples
have been interpreted \cite{Her,Ber,BT} as caustics of luminous matter
from a small galaxy that was ``eaten" by the giant elliptical.  Computer
simulations of the infall of the small galaxy in the fixed gravitational
potential of the giant elliptical show that the small galaxy gets tidally
disrupted, and that its stars end up on a thin ribbon in phase space.  
This ribbon is similar to the phase space sheet described earlier except
that it is more limited in spatial extent.  The ripples in the
distribution of light around giant ellipticals are the outer caustics
caused by the folding of the ribbon of stars in phase space.  As far as
we know, there is no other successful explanation of the ripples.

The appearance of the ripples provides an existence proof of discrete
flows and caustics. Although made of ordinary matter, stars are
collisionless when they fall in and out of a galactic gravitational
potential well.  That caustics occur in the distribution of stars 
answers the two concerns discussed above about the existence of discrete
flows and caustics of dark matter.  Indeed we may conclude from the
ripples' existence that 1) the velocity dispersion of a small galaxy 
is insufficient to erase caustics, and 2) flows do not get diffused by
scattering of inhomogeneities.  In particular, the dark matter is not 
in clumps of mass $10^{10}~M_\odot$ or larger because otherwise those
clumps would diffuse the flows of stars that cause ripples around giant
elliptical galaxies.

The ripples also answer other concerns that one may have with 
regard to discrete flows and caustics.  One concern sometimes
expressed is that the flows are unstable.   We do not know 
of a reason why the flows would be unstable.  But at any rate, 
the occurence of the ripples proves that the flows are not 
unstable.

\section{Further remarks}

Our large neighbor galaxy M31 is presently at a distance of approximately
730 kpc from us and has a line of sight velocity component in the Milky
Way rest frame of order 120 km/s in our direction \cite{BT}.  What role
does M31 play?  In the future, 5 Gyr from now say, the halos of M31 and
the Milky Way will be passing through each other.  The gravitational
potential will be more sharply time dependent then than it has been in the
past.  There will however continue to be discrete flows and caustics.  
Indeed, as was already emphasized, the existence of discrete flows and
caustics follows from just the assumption of cold dark matter.  The
distribution of caustics in both galaxies will be very much altered by the
encounter, but the caustics will not disappear.  Once M31 and the Milky
Way have joined into one big entity, their joint gravitational potential
well will cause a new inflow of surrounding dark matter, and a fresh
pattern of discrete flows and caustics will be established.  With regard
to the flows that exist in the Solar neighborhood today, M31's influence
is small because the particles that form those flows fell onto the Galaxy
a long time ago, at $t < t_0/2$, when M31 was much further from us than it
is now.  M31 does of course have an influence on the first one or three
Milky Way flows in its neighborhood, but this is a relatively small
fraction of the total phase space structure of the Milky Way halo.

A large part of the accepted lore on the structure of dark matter halos 
is based on the results of N-body simulations.  Are discrete flows and 
caustics seen in the simulations?  Actually, the existence and relevance 
of discrete flows and caustics are disputed on the basis of N-body 
simulations in refs. \cite{Hel,Moo}.  Still, for the reasons stated in 
the previous sections, discrete flows and caustics must be present in 
the simulations when the latter have adequate resolution. Ref. \cite{Moo}
argues that caustics have negligible relevance.  Yet, the velocity
distribution of dark matter particles in the simulation described
in that paper shows peaks (Figure 6 of \cite{Moo}).  Such peaks mean 
that there are discrete flows in the simulated halo.  And, if there are
discrete flows, caustics are present as well.

It should be emphasized that the resolution of present simulations is far
inadequate to describe the phase space structure of the Milky Way halo
down to our position in it.  Indeed, we found above that the minimum
number of flows at our location is of order 100.  Since phase space is 
6 dimensional, the minimum number of particles required to describe the
phase space structure of the Milky Way halo, down to our position, is
$100^6~=~10^{12}$.  Present simulations have only of order $10^7$ to
$10^8$ particles per galactic halo.  Also, because the particles in the
simulations are few they have to be proportionately heavy.  Each has a
mass of order $10^6~M_\odot$.  As a result, the simulations are afflicted
by two-body relaxation \cite{2bo}.  The simulated particles make hard
two-body collisions with one another, whereas two-body collisions between
axions or WIMPs are negligible.

Finally we note that discrete flows are evident in the N-body 
simulations of Stiff and Widrow \cite{Sti} who use a special 
technique to increase the resolution in the relevant regions 
of phase space.   Caustics are seen in the simulations of 
refs. \cite{Dor} and, more recently, ref. \cite{Shi}.

\section{Implications for Observation and Experiment}

In this section, we briefly list the implications of discrete flows 
and caustics for observation and experiment and refer to the growing
literature on this topic. 

According to the arguments of the previous sections, the velocity spectrum
of dark matter particles on Earth has at least twenty peaks due to flows of
particles that have fallen in and out of the Galaxy ten times or less in the
past.  The discrete flows have distinct signatures in direct dark matter
detectors on Earth.  Each flow produces a peak in the spectrum of microwave
photons from axion to photon conversion in cavity detectors of dark matter
axions \cite{duf}, and a plateau in the recoil energy spectrum of nuclei
struck by WIMPs in WIMP detectors \cite{Wimp}.  As a result of the orbital
motion of the Earth around the Sun, each of these spectral features has a
distinct annual modulation that depends on the velocity vector of the flow.  
If any of the direct dark matter detectors \cite{dama,cdms,edel,zepl} finds
a signal, it can in principle provide a wealth of information about the
structure of the Milky Way halo by observing the spectral features caused by
discrete flows.

If the Sun is close to a caustic, the flows that form that caustic 
have very large densities on Earth and hence produce prominent signals
in the direct dark matter detectors.   Although they have only received
bare mention in this paper, there are inner caustics \cite{stw,crdm,sin,nat} 
in galactic halos in addition to the outer caustics which we did discuss at 
some length.  Because the inner caustics are located relatively close to the 
galactic center, it is not unlikely that the Sun is near an inner caustic
of the Milky Way halo.

Although made of particles which move with typical halo velocities 
($v \sim 10^{-3} c$), the caustics themselves move much more slowly. 
The positions of caustics in galactic halos change only on cosmological 
time scales.  As a result, the caustics can accrete baryonic matter.  
The positions of the caustics in the halo may be revealed in this way
\cite{mil}.  Caustics can also produce bumps in galactic rotation
curves.  There is evidence that the galactocentric radii at which 
bumps occur in the rotation curves of external galaxies \cite{kin} 
and in the rotation curve of the Milky Way \cite{crdm,mil} are 
distributed in the same way as the radii of inner caustic rings in 
the self-similar halo model \cite{stw}.

WIMP annihilation is enhanced by the presence of caustics \cite{ann}
because the annihilation rate per unit time and unit volume is
proportional to the WIMP density squared.  If photons from WIMP
annihilation are observed the positions of caustics may be revealed 
as sharp lines and hot spots on the sky.

Finally, dark matter caustics produce gravitational lensing of distant
sources \cite{len}.  Because caustics have distinct density profiles, 
they have distinct gravitational lensing signatures as well.

\section{Acknowledgements}

This work was supported in part by the U.S. Department of
Energy under grant DE-FG02-97ER41029.  A.N. was supported 
in part by a grant from the J. Michael Harris Foundation.
P.S. thanks the Aspen Center for Physics for its hospitality 
while working on this paper.

\end{document}